\documentclass[aps,groupedaddress]{revtex4}

\usepackage[xdvi]{graphics}

\begin{document}

\title{Ground state numerical study of the
three-dimensional random field Ising model }
\author {I. Dukovski}
\affiliation{Department of Polymer Science and Engineering,
University of Massachusetts, Amherst, Massachusetts 01003}
\author {J. Machta}
\email{machta@physics.umass.edu}
\affiliation{Department of Physics,
University of Massachusetts, Amherst, Massachusetts 01003}
\date{\today}

\begin{abstract}   
The random field Ising model in three dimensions
with Gaussian random fields
is studied at zero temperature for system sizes up to $60^3$.
For each realization of the normalized random fields, the strength of the
random field, $\Delta$ and a uniform external, $H$ is adjusted to find the
finite-size critical point.  The finite-size critical point is
identified as the point in the $H$--$\Delta$ plane where three degenerate
ground states have the largest discontinuities in the
magnetization. The discontinuities in the magnetization and bond energy between
these ground states are used to calculate the magnetization and specific
heat critical exponents and both exponents are found to be near zero.
 
\end{abstract}

\maketitle

\section{Introduction}
\label{rfim}

The random field Ising model (RFIM) is among the simplest statistical mechanical models with quenched
disorder  but is still not well understood. It is presumed to describe equilibrium phase transitions in
physical systems such as fluids adsorbed in porous media and  diluted antiferromagnets.  However,
comparisons between theoretical predictions and experiments have been inconclusive because of the
difficulty of equilibrating the experimental systems.  For the three-dimensional RFIM, it is known that
there is an ordered phase for sufficiently low temperature and weak
randomness~\cite{ImMa75,Imbrie84,BrKu87}.  The standard picture~\cite{Villain85,Fish86,BrMo,Natt97} is
that the phase transition is continuous and is controlled by a zero
temperature fixed point with three scaling exponents.  The zero temperature (strong
disorder) fixed point implies that controlled renormalization group calculations cannot be carried
out so that information about exponents has come from numerical simulations, series
analysis~\cite{HoKhSe,GoAdAhHaSc} and real space~\cite{CaMa,FaBeMc} and other approximate
renormalization group calculations~\cite{GrMa82}.  There have also been suggestions that the transition is
first-order~\cite{HoKhSe,YoNa,BrDe98,MaNeCh00}  and, in fact, it is difficult to determine whether the
magnetization  vanishes continuously or discontinuously at the transition because the value of the
magnetic exponent, $\beta/\nu$ is very small.

Monte Carlo simulations~\cite{RiYo,Rieg95,NeBa,MaNeCh00} of the RFIM
suffer from long equilibration times and have been
limited to small systems.  The validity of obtaining critical
exponents from small systems has been called into question by simulations~\cite{MaNeCh00} showing
that for 
$24^3$ systems even qualitative features such as the apparent order of the transition vary from realization
to realization.  The difficulties of long equilibration times and small system sizes for Monte Carlo
simulations have prompted a number of studies of the zero temperature
RFIM~\cite{Ogielski,SwBrMaCiBa,HaNo,Sour98,HaYo01,MiFi02}.  Ground states of the RFIM can be
determined efficiently by mapping to the maximum flow problem and  then using a polynomial time
algorithm to solve the latter~\cite{HaRi01}.  The assumption of these studies is that the zero temperature
transition is in the same universality class as the  transition at nonzero temperature.  In this paper we
consider the zero temperature RFIM phase transition.

Estimates of many of the critical exponents for the three-dimensional RFIM are converging 
but there is still a problem with the specific heat exponent, $\alpha$.  Monte Carlo
simulations~\cite{Rieg95} and some zero temperature studies~\cite{HaYo01,NoUsEs} find $\alpha$ to be
quite negative, for example Hartmann and Young~\cite{HaYo01} find $\alpha=-0.65$.  On the other hand, a recent zero
temperature study by Middleton and Fisher~\cite{MiFi02} concludes that
$\alpha$ is near zero.  
Some experimental measurements of the specific heat~\cite{Wong86,SaKaToKa,BiFeHaHiRaTh}
show no divergence and can be interpreted as 
$\alpha$  near $-1$ while other measurements~\cite{BeKiJaCa,SlBe98} yield
$\alpha$ near zero and the experimental picture remains controversial~\cite{Wong96, BiFeHaHiRaTh,BeKlMo96}.  Large negative
values for $\alpha$ are in disagreement with the modified hyperscaling relation,
$\alpha=2-(d-\theta)\nu$, that is a central feature of the zero temperature fixed point picture. The
 relatively well established results that $\nu$ is in the range $1.1-1.4$ and that  $\theta$ is very close to $3/2$ imply that $\alpha$ is not much less than zero.  Very negative values of $\alpha$ are also inconsistent with the Rushbrooke relation $\alpha+2\beta +\gamma=2$ since
$\gamma$ is believed to be close to 2.

In this paper we study the zero temperature phase transition of the three-dimensional RFIM with
Gaussian random fields.  Our two primary goals are to provide evidence that the transition is,
indeed, continuous and to measure the specific heat exponent, $\alpha$.   The novel feature of our approach is that for each realization of the normalized random fields
we fine-tune both the strength of the random field and a uniform external field in order to bring the
system to its finite-size ``critical point'' (we refer to it as a critical point in anticipation of our result that the
transition is continuous).  Machta, Newman and Chayes~\cite{MaNeCh00} implemented a similar idea in
their Monte Carlo simulations.  We identify the finite-size critical point as the point where three degenerate
spin configurations have the largest jumps in the magnetization.  Critical exponents are extracted from the
finite-size scaling of the discontinuities in the magnetization and energy at the finite-size critical point.  Our results support the view that the
transition is continuous with
$\alpha$ and $\beta$ both near
zero.

\section{The random field Ising model at zero temperature}
\label{rfimatz}

The random field Ising model considered here is defined  
by the Hamiltonian:
\begin{equation}
\label{eq:def}
{\cal H}=-J\sum_{<i,j>}s_i s_j -\Delta \sum_{i} h_i s_i - H \sum_{i}s_i
\end{equation}
where $J$ is the coupling strength, $\Delta$ is the strength of the random field, $H$ is the uniform
external field,
$h_i$ is the normalized random field at site $i$ and $s_i$ is the
Ising spin variable at site $i$. $<i,j>$ indicates a sum
over nearest neighbor pairs on a three dimensional cubic lattice of linear size $L$ with periodic boundary conditions.
We take $J=1$ and the random fields to be Gaussian distributed with zero mean and unit variance,
\begin{equation}
P(h_i)=\frac{1}{\sqrt{2\pi}}\exp(-\frac{h^2_i}{2}).
\end{equation}
The two quantities of primary quantities that we measure are the magetization, $m$,
\begin{equation}
\label{eq:defm}
m=\frac{1}{L^3}\sum_i s_i 
\end{equation}
and the bond energy, $e$,
\begin{equation}
\label{eq:defe}
e=-\frac{1}{L^3}\sum_{<i,j>}s_is_j. 
\end{equation}

\begin{figure}
\begin{picture}(260,300)(0,0)
\put(20,20){\vector(0,1){240}}
\put(20,20){\vector(1,0){280}}
\put(0,230){$\Delta$}
\put(270,0){$T$}
\put(0,200){$\Delta_c$}
\put(240,0){$T_c$}
\put(100,100){$F$}
\put(220,220){$P$}
\put(0,0){\qbezier(20,200)(230,190)(240,20)}
\end{picture}
\caption{Phase diagram of the random field Ising model. The ordered ferromagnetic
phase is labeled $F$ and the disordered paramagnetic phase is labeled $P$.
The curve is a line of phase transitions.} 
\label{phase}
\end{figure}
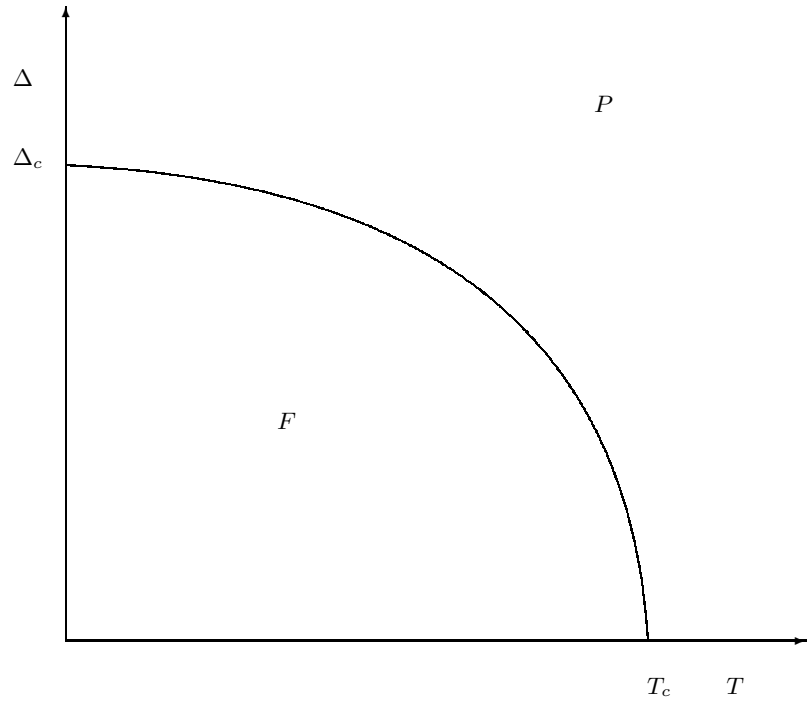

The presumed phase diagram of the three-dimensional RFIM is shown in 
Fig.~\ref{phase}.  The solid line is the phase transition between the ordered and disordered
phases. The  point
$(T_c,0)$ is the critical point of the pure Ising model while the
zero temperature phase transition is at the point $(0,\Delta_c)$. Assuming the absence of  special
points along the phase transition line (e.g.\ a tricritical point) 
the universal properties along the entire phase transition line, except at $\Delta=0$, are expected
to be the same as at the zero temperature transition.

\begin{figure}
\includegraphics{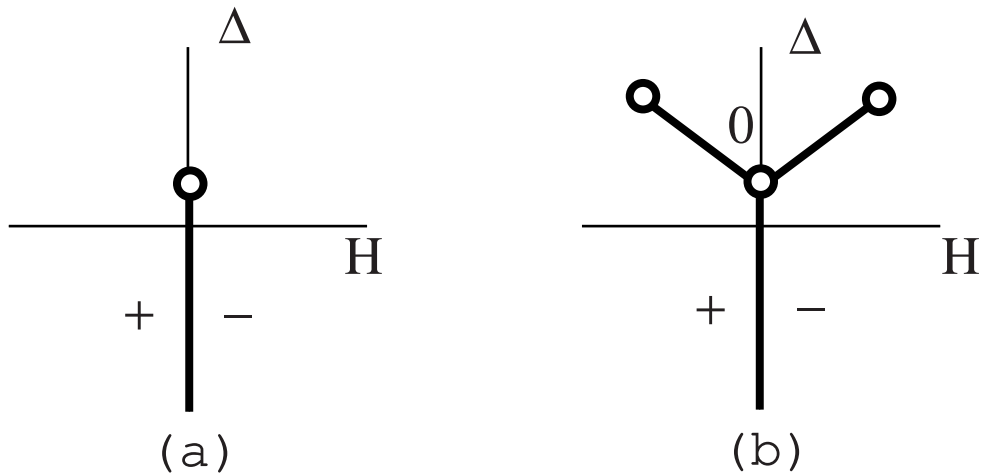}
\caption{Possible phase diagrams for the zero temperature phase transition:
 (a) continuous transition, (b) first-order 
transition. $+$ and $-$ 
are the coexisting ordered phases and $0$  is the coexisting disordered phase in (b). }
\label{ord}
\end{figure}

Consider the zero temperature transition. If $\Delta < \Delta_c$ the system is in one of two
ordered phases  so the magnetization as a function of $H$ has a jump at $H = 0$ while for $\Delta >
\Delta_c$, the magnetization is a continuous function of $H$.  If the transition is continuous, the
spontaneous magnetization, $\psi$ is expected to vanish as a power law as $\Delta$ approaches $\Delta_c$
from below,
\begin{equation}
 \psi\sim (\Delta_c-\Delta)^\beta
\end{equation}
where $\psi=m(H\rightarrow0^+)$ and $\beta$ is the magnetization exponent.
Figure \ref{ord}a illustrates the continuous transition scenario in the zero temperature $H$-$\Delta$
plane with a critical point at the end of a line of first-order transitions.
Another possibility is that the zero temperature transition is first-order.  A possible scenario is illustrated
with Fig.~\ref{ord}b. Here $(H=0,\Delta=\Delta_c)$ is a point of coexistence of two ordered phases
and one disordered phase.   The magnetization of the coexisting ordered phases is nonzero while the
disordered phase has zero magnetization. There will also be a nonzero ``latent heat" at the transition. 
Although entropy is ill-defined at zero temperature, it is reasonable to define latent heat in terms
of a discontinuity in the bond energy, $e$, between the ordered and disordered phases.

\begin{figure}
\includegraphics{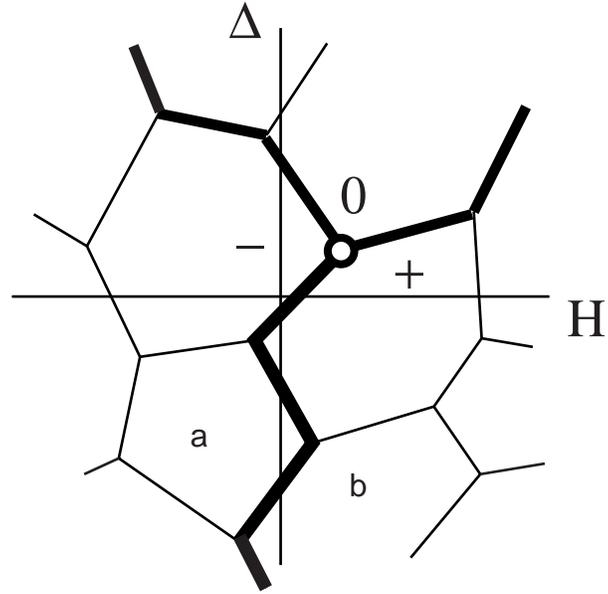}
\caption{Ground states 
for a given realization of normalized random fields, $\{h_i\}$.  
Coexistence lines between ground states with very different magnetizations are shown as thick
lines on the plot, the finite-size critical point is shown as an open circle and the three coexisting states at
the finite-size critical point are labeled $+$, $-$ and $0$  as described in the text.}
\label{coegdet}
\end{figure}

The foregoing applies to infinite systems.  The ground states of a typical finite system for a given
realization of the  random field 
are shown in Fig.~\ref{coegdet}.
Each point in the $H$-$\Delta$ plane corresponds to a single ground state 
of the system.  The set of points
corresponding to  a single ground state form a polygon 
since, for a given spin configuration, the energy is linear in both $H$ and 
$\Delta$.  In principle, there might be several degenerate ground states in a nonzero area of the
$H$-$\Delta$ plane but, for a continuous distribution of random fields, the probability of exact
degeneracy vanishes except along lines and at points.   Along the
edges between polygons the two ground states corresponding to each polygon are degenerate. Three ground
states are degenerate at ``triple points'' where three edges meet. In principle, more than three
ground states can be degenerate at a point but, for a continuous distribution of random fields, the
probability of more than three degenerate ground states vanishes.  

Degenerate ground states typically differ on a small fraction of spins but, corresponding to
the first-order line of the infinite system, some degenerate ground states differ by a large
fraction of the total number of spins.  The bold lines in Fig.~\ref{coegdet} correspond to large
jumps in the magnetization.  For $\Delta < \Delta_c$, there is a single jump between ground
states with large positive and negative magnetization and the separation between these ``phases'' is
the piecewise linear curve
$H_{\rm coex}(\Delta)$. This coexistence line is close to but not coincident with the
$\Delta$ axis.  The spontaneous magnetization,
$\psi$ is the magnetization of the positively magnetized ground states along $H_{\rm coex}$.  As
$\Delta_c$ is approached from below, $\psi$ decreases in steps at triple points.  Since the net
change in magnetization around a triple point is zero, the decrease in magnetization along $H_{\rm coex}$
is exactly the jump across the third edge or ``lateral line'' that defines the triple point.  As
the critical point is approached, the magnetization drops more precipitously so that the
magnetization jump across the lateral line increases.  We identify the finite-size ``critical point''
as the triple point with the largest jump across the lateral line.  In Fig.~\ref{coegdet} the finite-size
critical point is shown as an open circle and the corresponding lateral line is also shown as thick
line.  The three states that are degenerate at the critical point are called the $+$, $-$ and $0$ states
according to their relative magnetizations.  If the transition were first-order these states would be
the zero temperature analog of $+$ and $-$ ordered phases and the disordered phase that coexist at a
thermal first-order transition.  

One way to measure the discontinuity in the magnetization at the critical point is
via the quantity,
$m^\ast$,
\begin{equation}
m^{\ast}=\frac{1}{2}(m_+-m_0)(m_--m_0)(m_+-m_-)
\end{equation}
where  $m_+$, $m_0$ and $m_-$ are the magnetizations of the three 
coexisting critical ground states.  Note that the same combination of three magnetizations is small at triple
points  away from the critical point, even along the first-order line.

Although the energy, ${\cal H}$, is a continuous function of $H$ and $\Delta$, the bond energy, $e$,
has discontinuities across edges separating coexisting ground states. The bond energy along the
coexistence line, $H_{\rm coex}(\Delta)$, is expected to increase monotonically as $\Delta$ increases with jumps
across each lateral line.  The largest discontinuity in the bond energy is expected at the finite-size critical point and
the quantity $e^\ast$, is a measure of that discontinuity,
\begin{equation}
e^{\ast}=\frac{1}{2}(e_++e_-)(e_0-e_+)(e_0-e_-)
\end{equation}
where $e_+$, $e_-$ and $e_0$ are the bond energies of the  $+$, $-$ and $0$ states, respectively.  

If the phase transition is continuous, both $m^\ast$ and $e^\ast$ must approach zero as the system
size goes to infinity while if the transition is first-order, these quantities will saturate at
nonzero values.  Furthermore, if the transition is continuous, we propose the following finite-size
scaling behavior for the disorder averages of these quantities,
\begin{equation}
\label{eq:fssm}
\overline{m^\ast} \sim L^{-3\beta/\nu}
\end{equation} 
and
\begin{equation}
\label{eq:fsse}
\overline{e^\ast} \sim L^{-2(1-\alpha)/\nu}
\end{equation} 
where $L$ is the linear size of the system, $\alpha$ is the specific heat exponent, $\beta$ is the
magnetization exponent and $\nu$ is the  correlation length critical exponent.  The finite-size
scaling hypothesis for the magnetization is essentially identical to the standard finite-size
scaling hypothesis except that the measurement is made at a point that is fine-tuned for the given
realization of disorder rather than at the infinite system size critical point.  The finite-size
scaling for the bond energy energy is more difficult to motivate.  First, at zero
temperature free energy and energy are identical and entropy and specific heat are not well-defined.
It is reasonable to suppose that the derivatives of energy with respect to $J$ at zero temperature
have the same singularities as the derivatives of the free energy with respect to temperature at nonzero
temperature.  Thus, from Eq.\ (\ref{eq:def}), bond energy plays the role of entropy and its
derivative with respect to $J$ plays the role of specific heat.  The conventional finite-size scaling hypothesis for the specific heat is 
\begin{equation}
\label{eq:fssc}
C(T) \sim L^{\alpha/\nu} \tilde{C}((T-T_c)L^{1/\nu})
\end{equation}
If the specific heat is integrated across the finite-size rounding region defined by $|T-T_c| < L^{-1/\nu}$ we
obtain the finite-size ``latent heat'', $l$ of the transition, 
\begin{equation}
\label{eq:fssl}
l \sim L^{-(1-\alpha/)\nu} 
\end{equation}
At zero temperature, this latent heat is replaced by the discontinuity in the the bond energy.  Furthermore, since the bond energy changes at a discrete set of discontinuities, we
propose that a finite fraction of the latent heat is concentrated at the triple point with the
largest discontinuity, i.e.\ the point we have identified as the finite-size critical point.  Since
$e^\ast$ measures the square of the latent heat at the finite-size critical point, we obtain Eq.\
(\ref{eq:fsse}).

\section{Simulation method}

The main goal of our simulations is to find and measure the properties of the zero temperature,
finite-size critical point.  The finite-size critical point is  defined as the triple point with the largest jump in the magnetization between all
three coexisting states.   To find it we carry out an iterative search using a method similar to the
one described in Ref.\  \cite{FrGoOrVi}.  First, a point in the ordered phase on the first-order line  is
located.  For example, in Fig.\ ~\ref{coegdet} we locate ground states ``a'' and ``b'' and a point on the
coexistence line between them.  Next, the first-order line is followed in the direction of increasing $\Delta$
from one triple point to the next.  In the example of Fig.\ ~\ref{coegdet} the triple point where ``a'', ``b ''and ``+''
are degenerate is  located first.   At each triple point, the continuation of the first-order line (in this
example, between ``a'' and ``+'') and the lateral line  (between ``b'' and ``+'' ) are identified according to the relative
sizes of the discontinuities in the magnetization across the lines.  The sequence of triple points is recorded
and the triple point with the largest discontinuity in the magnetization across the lateral line is identified as the
finite-size critical point.

The most difficult computational problem in carrying out this program is to  find  the ground
state for given values of
$\Delta$, $H$ and $\{h_i\}$.  Finding ground states can be reduced to the problem of finding the maximum
flow on a graph 
\cite{Ogielski,AnPrRa} for which there exists polynomial time algorithms.
We use the push-relabel network flow algorithm \cite{GoTa88,GoCh97}, implemented in \cite{Go} as
version hi\_pr, to  find ground states.

Let's now describe in detail the two subroutines of the search algorithm.  Subroutine 1 finds a point on
the first-order line separating the positively and negatively ordered phases.   Two  ground states are found
deep in the ordered phases at values $\Delta_0=2.2$  and  $H=\pm0.1$. Once these ground states are
found, the algorithm calculates the value of $H$ where these two states are degenerate along the
$\Delta=\Delta_0$ line.  This point of degeneracy is easily found by the solution of two simultaneous linear
equations since once ${\cal H}$, $e$ and $m$ are known for a given spin configuration at some point in the
$H-\Delta$ plane, the value of
${\cal H}$ for that spin configuration at any other point is linearly related.  At this point of degeneracy the
ground state is calculated.  If the result is either of the two original ground states then we have found a
point on the first-order line.  If the ground state at the point of degeneracy is not one of the original states
then a new point of degeneracy is found between this new state and one of the  original two states. The
choice of which original state to pick is made by the  criterion that the jump in the magnetization
between it and the new state is biggest.  This process is repeated until the first-order line is found.

Subroutine 2 follows the first-order line found in subroutine 1 until the first triple point is
found in the direction of increasing $\Delta$.  From the point  found by subroutine 1 on the first-order line
a new point is picked by moving in the direction of the first-order line and increasing
$\Delta$ by
$0.1$. On Fig.~\ref{coegdet} this new point is on an imagined
continuation of the thick line between
``a'' and
``b'' in the  direction of $+$.
An increase of $0.1$ in $\Delta$ has proved to be sufficiently large to extend to a new ground state for all
system sizes studied.  Subroutine 2 is similar to subroutine 1 except that original ground states are always
retained in the iteration and a sequence of new ground states with decreasing values of $\Delta$  are
generated.  At each step in the iteration, the point of degeneracy is found between the new and the old
states and the ground state is computed at that point.  If the ground state is one of already identified
states then the triple point has been found otherwise the iteration is repeat. Call this first triple point
$(H_1,\Delta_1)$.

The triple point $(H_1,\Delta_1)$ defines three lines of coexistence, one of these lines is the previously
identified first-order line.  The jumps in magnetization across the other two lines are measured.  The line
with the smaller jump in magnetization is called the ``lateral line'' and the remaining line with the large
jump in magnetization is continuation of the first-order line.   The jump across the lateral line is recorded
and subroutine 2 is invoked again to find the next triple point along the first-order line.  By repeating this
set of steps many times, a sequence of triple points $(H_i,\Delta_i)$ are identified.  The triple point with
the largest jump across the lateral line is identified as the finite-size critical point, $(H^\ast,\Delta^\ast)$.

We note that while the critical point is identified by the size of the jumps in the magnetization, in every
case that we examined it also has the largest discontinuity in the bond energy. In principle, a
definition of the finite-size critical point based on the bond energy discontinuity might sometimes yield a
different triple point.

A typical running time for finding the ground state for a system of $L=10$ 
is $6\times10^{-5}$ sec/spin on a Pentium III 750MHz machine. The total running time of the
entire algorithm for the same system size is approximately one minute.
We simulated systems of size up to $L=60$. The number of realizations of disorder range from 23 to
126 for different system sizes.  The small number of realizations are a consequence of the fact that many
ground states must be explored for each realization of disorder to find the finite-size critical point.

\section{Results}

\begin{figure}
\includegraphics{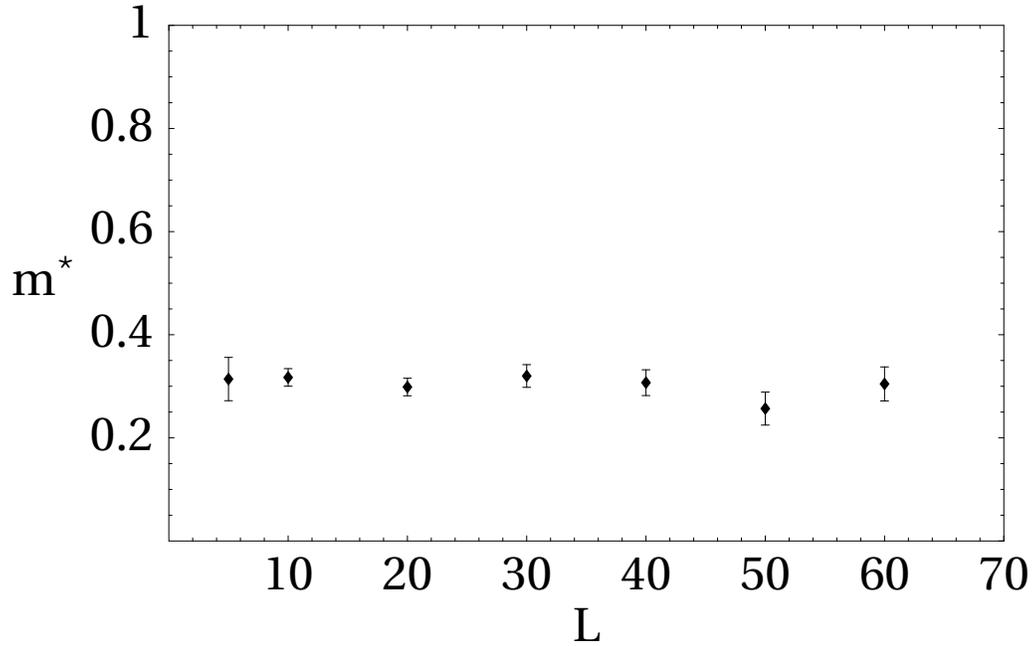}
\caption{The magnetization discontinuity at the critical point, $m^{\ast}$ vs.\ system size, $L$.}
\label{magn}
\end{figure}

\begin{figure}
\includegraphics{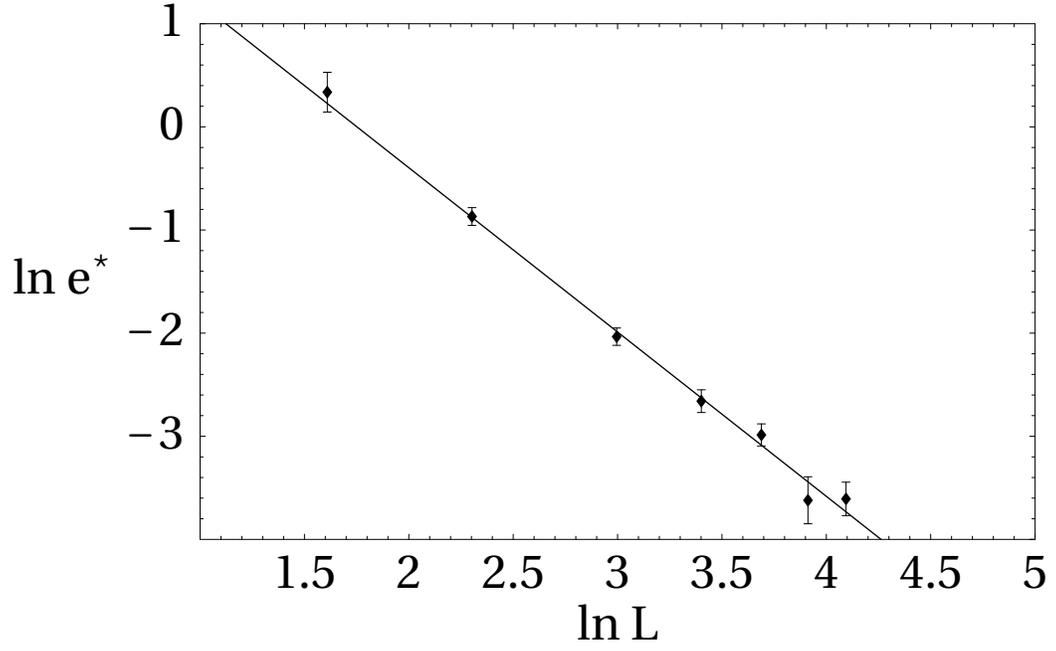}
\caption{Log-log plot of bond energy discontinuity at the critical point, $e^{\ast}$ vs.\ the system size, $L$. The solid line is a
fit as described in the text.}
\label{logenerg}
\end{figure}

Figure \ref{magn} shows the critical magnetization discontinuity  
$m^{\ast}$ as a function of system size $L$.
Clearly this plot, together with Eq.\ (\ref{eq:fssm}) suggests that the magnetization discontinuity  at
the critical point does not decrease with $L$ which can be interpreted
either as a first-order transition or a value of $\beta$ very near zero.  The quality of the data is
not sufficient to make a useful measurement of $\beta$.  Figure \ref{logenerg} shows the finite-size
scaling of the critical bond energy discontinuity $e^{\ast}$. A fit of the form:
\begin{equation}
e^{\ast}=a+bL^{-c}
\end{equation}
yields $a=0.007\pm 0.005$, $b=21\pm5$ and $c=1.7\pm0.1$.
It is clear that $e^{\ast}$ vanishes as power of $L$.
This result is a clear indication of a continuous transition.
Assuming a=0, we obtain a fit, shown in Fig.~\ref{logenerg}, of the form:
\begin{equation}
e^{\ast}=bL^{-c}
\end{equation}
with $b=16.0\pm2.6$ and $c=1.59\pm0.05$.  From the latter result and Eq.\ (\ref{eq:fsse}) we obtain
$(1-\alpha)/\nu=0.80\pm0.03$ (with $\chi^2=2.84$, 
$\chi^2/d.o.f.=0.57$ and $Q=0.73$).  This result is in agreement with Middleton and Fisher's
value\cite{MiFi02}, $0.82\pm0.02$.
 
\begin{figure}
\includegraphics{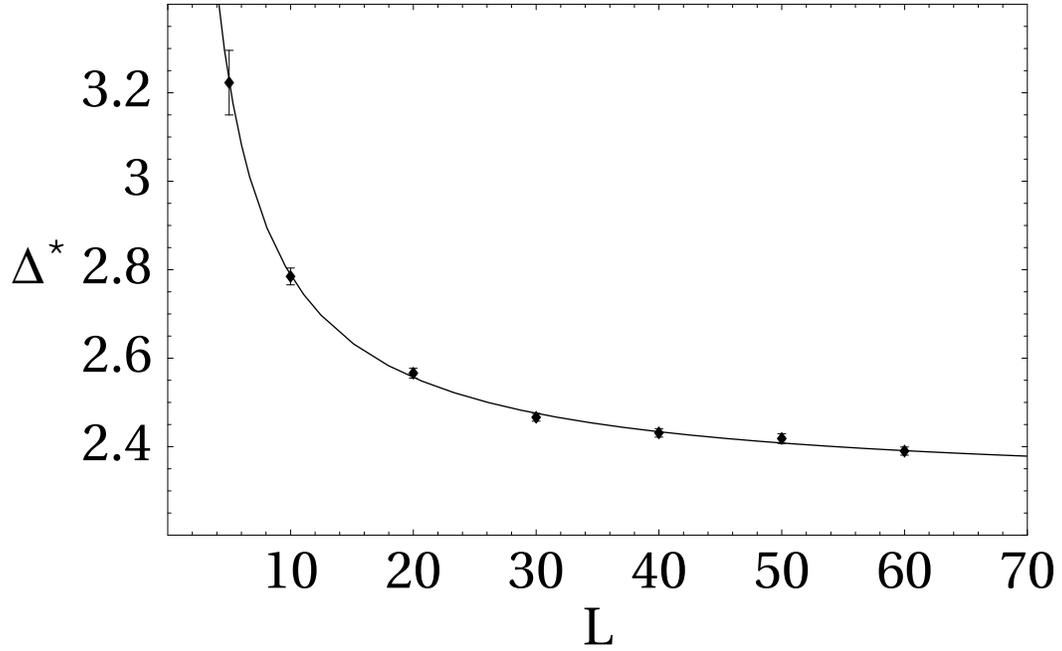}
\caption{The critical strength of randomness, $\Delta^{\ast}$ vs.\ system size, $L$. The solid line is a 
fit as described in the text.}
\label{delta}
\end{figure}

The correlation length exponent $\nu$ and the infinite size critical disorder strength, $\Delta_c$ 
can be obtained from the finite size scaling of $\Delta^{\ast}$. The fit in Fig.~\ref{delta}
is:
\begin{equation}
\Delta^{\ast}=\Delta_c+bL^{-1/\nu}
\end{equation}
with $\Delta_c=2.29\pm 0.02$, $b=4.1\pm0.6$ and $\nu=1.1\pm0.1$ (with $\chi^2=2.79$, 
$\chi^2/d.o.f.=0.69$ and $Q=0.59$).   
This result is in relatively good agreement with recent results in the literature 
as shown on Table \ref{tab1}.  The value of $\nu$ is somewhat smaller than the recent values in
Refs.\ \cite{HaYo01,MiFi02} which may result from the use of a smaller maximum system size in
our study.  Combining our results for $\nu$ and for $(1-\alpha)/\nu$ yields $\alpha=0.1\pm0.1$. 
Using the larger values of $\nu$ from Refs.\ \cite{HaYo01,MiFi02} gives slightly negative values
of $\alpha$.  In any case, within the uncertainties, our results are consistent with modified
hyperscaling.

\section{Discussion}

The main result of this study is a new measurement of the specific heat exponent of the RFIM.  Our result
is a value of $\alpha$ near zero that is consistent with the modified hyperscaling relation and the recent
simulation results by Middleton and Fisher~\cite{MiFi02}.  It is not clear why a very similar measurement
by Hartmann and Young~\cite{HaYo01} give $\alpha$ quite negative.  Our value of $\alpha$ and that of Ref.\ \cite{MiFi02} are based directly on the finite-size scaling of the bond energy at the critical point whereas in Ref.\  \cite{HaYo01} the average bond energy is numerically differentiated to obtain the specific heat
and the finite-size scaling of the peak is used to obtain $\alpha$.

\acknowledgements
This work was supported by NSF grants DMR-9978233.  We thank Alan Middleton, Po-zen Wong and David Belanger for useful comments.

\newpage

\printtables

\begin{center}
\begin{table}
\begin{tabular}{rcrcccrcr}
\hline
Ref.&\hspace{.1in}&$\Delta$&\hspace{.1in}&$\nu$
&\hspace{.1in}&$(1-\alpha)/\nu$&\hspace{.1in}&$\alpha$\\  
\hline  
This 
work&\hspace{.1in}&2.29(2)&\hspace{.1in}&1.1(1)&\hspace{.1in}&0.80(3)&\hspace{.1in}&0.12\\
\cite{HaYo01}&\hspace{.1in}&2.28(1)&\hspace{.1in}&1.36(1)&\hspace{.1in}&1.20&
\hspace{.1in}&$-0.63(7)$\\
\cite{MiFi02}&\hspace{.1in}&2.270(4)&\hspace{.1in}&1.37(9)&\hspace{.1in}&0.82(2)&
\hspace{.1in}&$-0.12$\\
\cite{AnSo}&\hspace{.1in}&2.26(1)&\hspace{.1in}&1.22(6)\\
\cite{HaNo}&\hspace{.1in}&2.29(4)&\hspace{.1in}&1.19(8)\\
\cite{NoUsEs}&\hspace{.1in}&2.37(5)&\hspace{.1in}&1.0(1)&\hspace{.1in}&1.55&
\hspace{.1in}&$-0.55(20)$\\
\hline\\
\end{tabular}
\caption{A summary of recent zero temperature estimates of $\Delta_c$, $\nu$, $(1-\alpha)/\nu$, and
$\alpha$.  For $\alpha$ and
$(1-\alpha)/\nu$ the value without error estimated is derived from the other, directly measured, value
and the same authors' value of $\nu$.}
\label{tab1}
\end{table}
\end{center}

\newpage

\printfigures

\newpage

\newpage

\newpage

\newpage

\newpage

\newpage

\end{document}